\documentclass{mem}
\usepackage{natbib}
\bibpunct{(}{)}{;}{a}{}{,} 
\usepackage{txfonts}
\usepackage{balance}
\usepackage{graphicx}
\usepackage[a4paper]{hyperref}
\idline{0}{0}

\begin{document}

\title{Polarization diagnostics of proton beams\\ in solar flares}

\author{J.\,\v{S}t\v{e}p\'an\inst{1,2}}

\institute{
     LERMA, Observatoire de Paris -- Meudon, CNRS URA 8112, 5,
     place Jules Janssen, 92195 Meudon Cedex, France\\
     \email{Jiri.Stepan@obspm.fr}
     \and
     Astronomical Institute, Academy of Sciences of
     the Czech Republic, Fri\v{c}ova 298,
     25165 Ond\v{r}ejov, Czech Republic
}
 
\authorrunning{\v{S}t\v{e}p\'an}

\titlerunning{Polarization diagnostics of proton beams in solar flares}

\abstract{
We review the problem of proton beam bombardment of solar chromosphere
considering the self-consistent NLTE polarized radiation transfer in hydrogen
lines. Several observations indicate a linear polarization of 
the H$\alpha$ line
of the order of 5\% or higher and preferentially in radial direction.
This polarization is often explained as anisotropic collisional excitation
of the $n=3$ level by vertical proton beams. Our calculations indicate that
deceleration of the proton beam with initial power-law energy distribution
together with increased electron and proton densities in the
H$\alpha$ forming
layers lead to a negligible line polarization. Thus the proton beams seem
not to be a good candidate for explanation of the observed polarization degree.
On the other hand, the effect of electric return currents could perhaps
provide a better explanation of the observed linear polarization.
We report the new calculations of this effect.
}

\maketitle{}


\section{Introduction}

There is an observational evidence for the fast electron beams
(\mbox{10--100 keV}) and also for the fast proton beams ({}$\sim 10$~MeV)
bombarding the solar chromosphere during the impulsive and gradual phase of
solar flares \citep{korchak67,orrall76}.
This bombardment is
consistent with the ``standard model'' of solar flare which assumes an
injection of the high energetic particles from a coronal reconnection site
to the chromosphere. It results in heating and nonthermal excitation of
the chromospheric gas.
The standard semiempirical models of the flaring chromosphere
\citep{machado80}
are based
on the series of continuum and lines observations and do not
take into account a nonthermal excitation; they rather
overestimates the chromospheric temperature to explain
an increased radiative emission.

Also the low energy proton beams (below 1~MeV)
could play a significant role in the
flare physics but their presence in the chromospheric layers is still
uncertain due to their negligible bremsstrahlung radiation. There are however
different ways how to diagnose them. For instance:
\begin{enumerate}
\item Ly$\alpha$ red wing emission due to charge exchange effect
\citep{cc85,zhao98}.
\item Modification of the lines intensity profiles due to nonthermal
excitation \citep{henoux93,xu05b}.
\item Detection of a linear polarization of the lines due to an
anisotropic collisional excitation (i.e. impact atomic polarization).
\end{enumerate}
Moreover, several unanswered questions still remain about the electron beams.
Especially the physics of the return currents creation and
their influence on lines formation.

Measurement of linear polarization in solar flares is a complicated task
due to the high spatial and time gradients.
Lot of observations report the polarization degree of H$\alpha$
of the order of few percent
\citep{henoux90b,vogt99,hanaoka03,xu05a}.
Direction of this polarization is usually found to be either radial
or tangential with respect to the limb
and its degree is up to 5\% or even exceeds 10\% in the few
cases.
The standard interpretation of these measurements is anisotropic
excitation by proton and/or electron beams.
By contrast,
no measurable linear polarization above the noise level at 0.1\%
was found in the set of many different flares by another authors
\citep{bianda03,bianda05}.

Our objective is to find out if the measurements of a linear polarization
degree of the hydrogen H$\alpha$ line are a promising tool for
the diagnostics of the low-energy proton beams.
Moreover, we present a first
quantitative estimation of the polarizing effect of the return electric
currents taking properly into account depolarizing collisions and radiative
transfer.

We start with the standard semiempirical model of the flaring
atmosphere F1 \citep{machado80} with a fixed temperature structure.
We include the nonthermal collisional rates
of excitation and ionization 
to obtain the new line profiles.
We use a stationary modeling of the polarized radiation transfer
in the chromospheric hydrogen lines using a multilevel
NLTE polarized transfer code.


\section{H$\alpha$ impact polarization}

\begin{figure}
\centering
\includegraphics[width=\columnwidth]{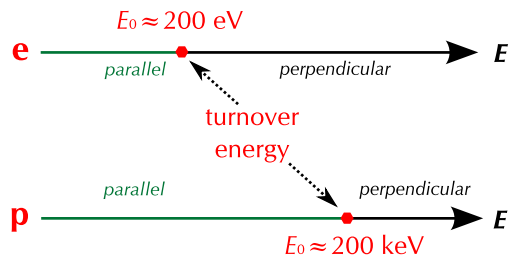}
\caption{\footnotesize
Linear polarization of the photons emitted in the H$\alpha$ line
perpendicularly to the collisional trajectory (i.e. in
the direction of a maximum polarization degree).
For both the electrons and protons impacts, low energies of collisions
lead to linear polarization of the emitted photons in the direction
parallel to the collisional trajectory. A polarization direction
above the so-called turnover energy is perpendicular to it.
}
\label{fig:turnover}
\end{figure}

Let us assume a beam of the unidirectional charged colliders and let 
their energy be high enough to excite the $n=3$ level
(i.e. higher than 12.1~eV).
In fact, the $n=3$ level is composed of five quasi-degenerated fine
structure levels. Some of them can be polarized, hence the photons
emitted in the H$\alpha$ (and Ly$\beta$) line are in general polarized.
An anisotropical excitation of these levels by either protons or electrons
is one of the processes which can lead to polarization
of these levels. The orientation of electric
vector of this radiation can be either parallel or perpendicular to
the velocity of the beam (cf. Figure~\ref{fig:turnover}).
A beam is expected to follow the direction
of magnetic field which is assumed to be vertical in the atmosphere;
hence the linear polarization is either radial or tangential
with respect to the solar limb.

\begin{figure}
\centering
\includegraphics[width=\columnwidth]{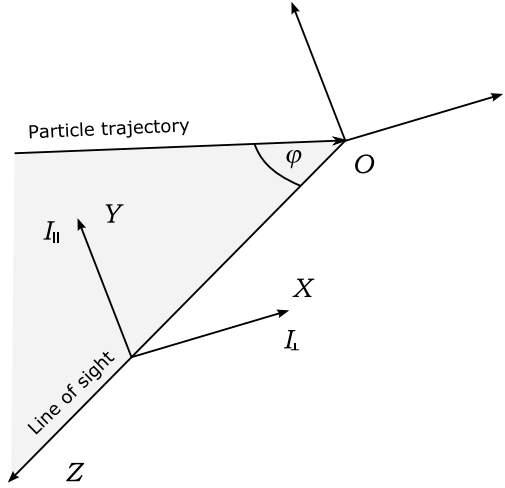}
\caption{\footnotesize
We use a right-handed coordinate system with $Y$-axis in the collisional
plane and $Z$-axis in the line of sight direction. Due to cylindrical symmetry
of the model, only Stokes parameters $I$ and $Q$ do not generally
vanish in this coordinate system.}
\label{fig:refframe}
\end{figure}

The assumptions above should lead to an upper theoretical limit for
the H$\alpha$ linear polarization. The coordinate system definition
is given in Figure~\ref{fig:refframe}. In such a reference frame,
only the Stokes parameters $I$ and $Q$ are in general nonzero.
We are interested especially in a modeling of the radial polarization.
It follows from
Figure~\ref{fig:turnover} that this orientation is associated with
the low-energy electrons or the low-energy protons moving
vertically in the atmosphere.\footnote{The case of horizontal motions
at high energies is unlikely for the physical reasons.}


\section{Formalism and methods}

A vertical magnetic field is assumed to be
of the strength of few hundreds gauss \citep{vogt97}.
It can be shown that
in this regime and
in the plasma conditions under consideration (see below)
fine structure splitting of the hydrogen levels
up to $n=3$ is a good approximation \citep{sahal96}. The hyperfine
splitting of the levels can be neglected because it does not significantly
affect a
linear polarization degree \citep{bommier86a}. 
In addition, a lifetime of the
fine structure levels is reduced by collisions with the
background protons and electrons and thus the hyperfine levels
of the excited states completely overlap.
To describe an atomic state we adopt the formalism of the atomic density
matrix in the basis of irreducible tensorial operators $\rho^K_Q(nlj)$
\citep{fano57}, with the traditional meaning of the symbols.
We neglect all the quantum coherences between
different fine structure levels due to their large separation in comparison
to their width and due to the selection rules for the optical transitions.
A magnetic field is able to destroy all the quantum coherences between the
Zeeman sublevels
of any level. In the formalism of irreducible tensors it means that
all the multipole components of $\rho$ with $Q\neq 0$ are identically
zero, $\rho^K_Q(nlj)=\delta_{Q0}\rho^K_Q(nlj)$. On the other hand,
the strength of a magnetic field is not so high to induce
a Zeeman splitting large enough to lead to the complicated level-crossing
effects. Finally, because of the assumption of absence of any
circularly polarized radiation, the odd ranks $K$ of the density matrix
are identically zero. All these assumptions lead to the simplifications of
the formalism.

The results of our modeling are strongly dependent on the collisional
cross-sections of the different excitation and ionization transitions.
We consider the following collisional processes and cross-sections
data:
\begin{itemize}
\item Excitation and charge exchange for a proton beam
using the close-coupling cross-sections data of \citet{balanca98}.
\item The dipolar transitions between the fine structure levels
$nlj\to n l\pm 1 j'$ for interaction with the background electrons,
protons, and the beam are calculated using the semiclassical theory
of \citet{sahal96}.
\item Excitation of the levels populations by the background electrons
(the cross-sections data are taken from the AMDIS database,
http://www-amdis.iaea.org).
\end{itemize}
For the purposes of impact polarization studies it is necessary to
calculate all the collisional transition rates
\begin{equation}
C^{K\to K'}_{nlj\to n'l'j'}=N_{\rm P}\int{\rm d}^3\vec{v}f(\vec{v})v
\sigma^{K\to K'}_{nlj\to n'l'j'}(\vec{v})\;,
\end{equation}
for transitions $\rho^K_0(nlj)\to\rho^{K'}_0(n'l'j')$.
These rates enter the equations of statistical equilibrium
together with the radiative rates $R^{K\to K'}_{nlj\to n'l'j'}$
in the so-called impact approximation
\citep{landi84,bommier91}.
The equations of statistical equilibrium
have the form\footnote{Additional terms have to be added to
take into account the bound-free transitions. They are not expressed in
the Eq.~(\ref{eq:ese}) for the reasons of simplicity.}
\begin{equation}
\sum_{nljK} (C^{K\to K'}_{nlj\to n'l'j'}+R^{K\to K'}_{nlj\to n'l'j'})
\rho^K_0(nlj)=0\;.
\label{eq:ese}
\end{equation}

To take properly into account a coupling of the atomic states
at the different atmospheric points, it is necessary to solve
a system of equations~(\ref{eq:ese}) coupled by
the radiative transfer equation for Stokes vector
$\vec{I}=(I,Q)^{\rm T}$. Its evolution along the
path $s$ is given by
\begin{equation}
\frac{{\rm d}\vec{I}}{{\rm d}s}
	=\vec{J}-\vec{K}\,\vec{I}\;.
	\label{eq:rte}
\end{equation}
In this equation, $\vec J$ is the emissivity vector and $\vec K$ is the
so-called propagation matrix.
We use our multigrid code \citep{stepan06spw}
to solve this problem in the plan-parallel geometry,
out of the LTE approximation (the case of the chromospheric
H$\alpha$ line).


\section{Results for the proton beams}

The temperature
structure of the atmosphere is fixed and given by the F1 model.
A new ionization degree and the atomic states along the chromosphere
are obtained by solution
of the system of the Eqs.~(\ref{eq:ese}) and (\ref{eq:rte}).
\begin{figure}
\centering
\includegraphics[width=\columnwidth]{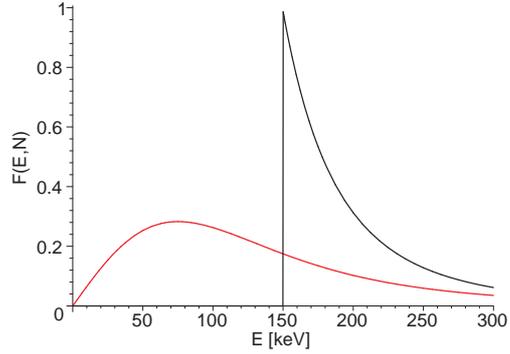}
\caption{\footnotesize
Energy distribution of a beam flux (in arbitrary units),
$\delta=4$, $E_{\rm c}=150~{\rm keV}$.
The initial sharp-peaked power-law energy distribution (upper curve) becomes
flatter in the region where the slowest protons are stopped (lower curve).
}
\label{fig:distrib}
\end{figure}
The initial energy distribution of the proton
beam at the top of the chromosphere
is (as usually) expected to be given by the power-law
\begin{displaymath}
F(E)\sim E^{-\delta}\;,\qquad E>E_{\rm c}\;,
\end{displaymath}
with the lower energy cut-off $E_{\rm c}$
and the spectral index $\delta$.
After crossing some column depth, the distribution is modified by
collisions with various atmospheric species (see Figure~\ref{fig:distrib}),
\citep{emslie78,cc85}.
The protons most efficient in creation of the $n=3$ level polarization
have a small energy about $5~{\rm keV}$ \citep{balanca98}.
A number of such protons in the region of the H$\alpha$ line core formation
is very small compared to the number of protons of an
energy about $E_{\rm c}$ at
the injection site \citep[cf.][]{vogt97,vogt01}.
As a result, there is only a small number of
low-energy protons even for the high initial beam fluxes.

\begin{figure}
\centering
\includegraphics[width=\columnwidth]{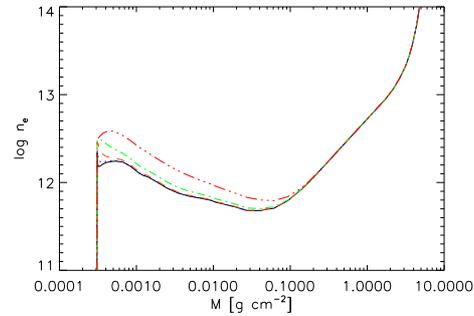}
\caption{\footnotesize
Electron density for the model F1. The plot
corresponds to the case of $\delta=4$, $E_{\rm c}=150\;{\rm keV}$.
Thermal case is plotted by a solid line,
the non-thermal beam fluxes are $\mathcal{E}_0=10^8$ (dot),
$10^9$ (dash), $10^{10}$ (dash-dot),
and $10^{11}\,{\rm erg\,cm^{-2}\,s^{-1}}$ (dash-dots).
}
\label{fig:fioniz}
\end{figure}

The electron (proton) densities of the chromosphere obtained as a result of the
NLTE calculations are of the order of $10^{12}~{\rm cm^{-3}}$ in the regions
of the H$\alpha$ centre and near wings formation
(Figure~\ref{fig:fioniz}). These densities were computed using the
code of \citet{kasparova02} adapted for the proton beams studies.
It was shown by \citet{bommier86b} that depolarization of the hydrogen
levels by collisions with the background electrons and protons
becomes important even at the densities of the order of
$10^{10}~{\rm cm^{-3}}$.
It is therefore expected that
the depolarization effect will play a crucial role
in the chromospheric conditions.

\begin{figure}
\centering
\includegraphics[width=\columnwidth]{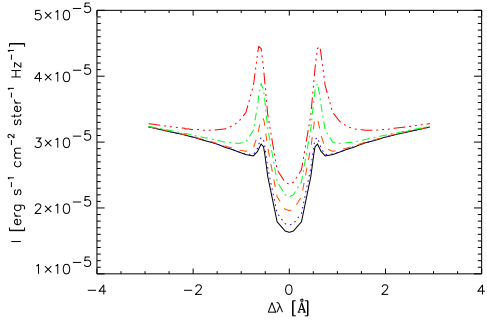}
\caption{\footnotesize
H$\alpha$ line intensity profiles for the same set of models
as in the Figure~\ref{fig:fioniz}.
}
\label{fig:ihalpha}
\end{figure}

\begin{figure}
\centering
\includegraphics[width=\columnwidth]{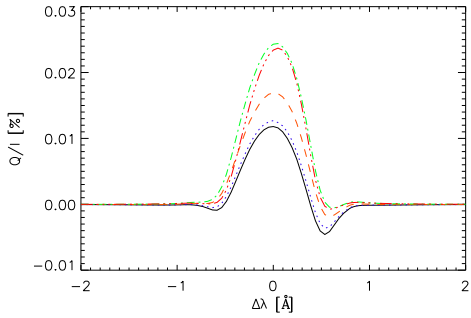}
\caption{\footnotesize
Emergent fractional linear polarization $Q/I$ profiles (\%)
computed close to the limb ($\mu=0.11$) for the same set of models
as in the Figure~\ref{fig:fioniz}.
A positive sign of $Q/I$ means
the tangential direction
of polarization, while a negative sign is the radial one.
}
\label{fig:qihalpha}
\end{figure}

The theoretical line profiles calculated for a set of the
proton beam energy fluxes are plotted in the Figures~\ref{fig:ihalpha}
and~\ref{fig:qihalpha}.
As it can be seen in Figure~\ref{fig:qihalpha}, a polarization degree
of the emergent H$\alpha$ line close to the limb is extremely small. Moreover,
its orientation is tangential and not radial as it was
expected for the low-energy
proton impacts. This polarization is mainly due to resonance scattering of
the radiation, while the impact polarization effect is not seen.
The degree of 0.02\% is well below any measurable value to day.


\section{Effect of electronic return currents}

It seems that linear polarization of the H$\alpha$ line due to slow
proton beams is not a good diagnostics tool even if very restrictive
conditions on the beam anisotropy are postulated. Another phenomena
which could explain the observed radial polarization are the
return currents (RC) associated with the electron beams.
The detailed studies of the
collective plasma processes show that
the return currents which neutralize a huge electric current of the beam
can be formed if several plasma conditions in the atmosphere are
fulfilled \citep{norman78}.
The atomic excitations caused by the return current
were shown to be more important than the effect of the beam itself
in some depths \citep{karlicky04}.

\citet{karlicky02} propose the anisotropical impacts of
the slow (few \mbox{deca-eV}) RC as an explanation for the observed 
linear polarization of the H$\alpha$ line.
However, all these calculations neglected the effects of depolarizing
collisions and polarized radiative transfer.
We have studied a simple model of monoenergetic electron beam with the
initial energy flux of $1.2\times 10^{12}~{\rm erg\;cm^{-2}\;s^{-1}}$
composed of the electrons with the energy 10~keV penetrating the
F1 atmosphere. The electron-hydrogen cross-sections for the dipolar
transitions have been calculated by the semiclassical method with
momentum transfer \citep{bommier05}.
These calculations show that return current is locally able to produce
a significant atomic polarization, but it is decreased by the
depolarizing collisions and high intensity of radiation.
If all these effects are taken into account together with radiation transfer,
the emergent linear polarization degree in the centre of H$\alpha$ is
only about 0.25\% in the radial direction.
That is still one order of magnitude below the measured values.

\section{Conclusions}
A net impact polarization of an electron beam + RC seems to be a more promising
explanation of the observed linear polarization than
the proton beam impacts.
However, collisional depolarization by background electrons and protons
together with the increased radiation intensity can
destroy most of the impact atomic polarization.

Several things have been simplified to make our models traceable:
\begin{enumerate}
\item The semiempirical model F1 overestimates the atmosphereric temperature.
As a result, depolarization and lines intensities are also overestimated.
\item No time dependence was considered. Detailed time-dependent modeling
could possibly explain some aspects of polarization generation.
\item The plane-parallel model has a limited applicability and more complex
geometry could be considered to interpret the observations.
\item If a RC is carried by a substantial number of background electrons,
they do not contribute to the collisional depolarization in the same way as
the thermal electrons. That was not taken fully
into account in our models.
\item The beams are never purely directional and vertical and
their scattering may lead to further decrease of the impact polarization.
The monoenergetic electron beam + RC is not a satisfactory physical model.
\end{enumerate}
A quantitative estimation of all these effects should be a subject of the
future studies.


\bibliographystyle{aa}
\bibliography{bibs}

\end{document}